\documentclass[twoside,fleqn]{ActaStyle}
\usepackage{times,amssymb}
\usepackage{texdraw}

\usepackage{lscape}
\usepackage{graphicx,epsfig}
\usepackage[tbtags]{amsmath}

\newcommand{\bphi}{\boldsymbol{\phi}}
\newcommand{\bx}{\boldsymbol{x}}
\newcommand{\bxp}{\boldsymbol{x}_\|}
\newcommand{\bxperp}{\boldsymbol{x}_\perp}
\newcommand{\e}{{\mathrm e}}
\newcommand{\Or}{{\mathrm O}}

\def\authorheading{H.~W.~Diehl}

\def\shorttitle{Critical behavior at $m$-axial Lifshitz points}

\begin{document}

\pagerange{1}{13}

\title{%
CRITICAL BEHAVIOR AT M-AXIAL LIFSHITZ POINTS\footnote{Based on a
  plenary talk given at the 5th International Conference
``Renormalization Group 2002'', Tatranska Strba, High Tatra Mountains,
Slovakia, March 10--16, 2002.}
}

\author{H.~W.~Diehl\email{phy300@theo-phys.uni-essen.de}}
{Fachbereich Physik, Universit{\"a}t Essen, D-45117 Essen, Federal
Republic of Germany}

\day{May 10, 2002}

\abstract{%
  An introduction to the theory of critical behavior at Lifshitz
  points is given, and the recent progress made in applying the
  field-theoretic renormalization group (RG) approach to $\phi^4$
  $n$-vector models representing universality classes of $m$-axial
  Lifshitz points is surveyed. The origins of the difficulties that
  had hindered a full two-loop RG analysis near the upper critical
  dimension for more than 20 years and produced long-standing
  contradictory $\epsilon$-expansion results are discussed. It is
  outlined how to cope with them. The pivotal role the considered
  class of continuum models might play in a systematic investigation
  of anisotropic scale invariance within the context of thermal
  equilibrium systems is emphasized. This could shed light on the
  question of whether anisotropic scale invariance implies an even
  larger invariance, as recently claimed in the literature.}

\section{Preliminary remarks}
\label{sec:intr} \setcounter{section}{1}\setcounter{equation}{0}

The aim of this article is to give a brief introduction to the field
of critical behavior at Lifshitz points \cite{Hor80,Sel92}, and to
survey the recent progress that has been made in applying the
field-theoretic renormalization group (RG) approach
\cite{MC98,MC99,DS00a,SD01,DS01a,DS02}. As is appropriate for a
conference on the topic `renormalization group', I shall assume some
basic familiarity of the reader with RG ideas. However, in view of the
mixed background of the participants of the conference, no extensive
knowledge of the relevant condensed matter physics is presupposed.

Since the literature on Lifshitz points---or, more generally, on
systems with spatially modulated phases---is vast, it is impossible to
mention or even cite all relevant papers. The paper is designed to
give a reasonably self-contained account of the issues on which we
focus, and to serve as a guide to the literature. The choice of the
references has been made accordingly. There exist extensive review
articles \cite{Hor80} and \cite{Sel92}, which survey the literature
till 1992 and contain extensive lists of references. The reader may
consult these for further information on topics that had to be left
out here.

\section{Introduction and background}
\label{sec:backg}

\subsection{Generic phase diagram with a Lifshitz point}
\label{sec:gpd}
\begin{figure}[htb]
\begin{center}
\includegraphics[width=8cm]{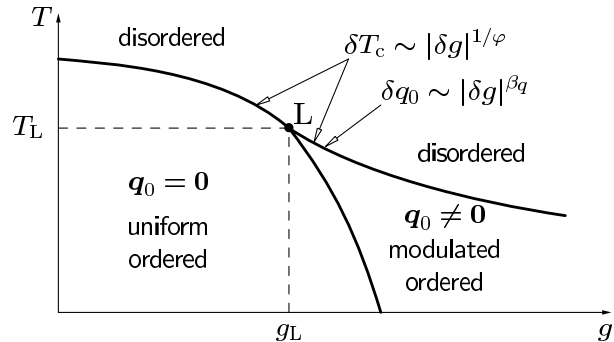}
\end{center}
\caption{Schematic phase diagram with a Lifshitz point L. The
  disordered phase is separated from the uniform and modulated ordered
  phases via a critical line. The crossover exponent $\varphi$ is
  defined via the behavior of the critical line near L; the
  wave-vector exponent $\beta_q$ describes how the modulation
  wave-vector $q_0$ tends to the value $q_{0,{\mathrm L}}$ as L is
  approached along the critical line between the disordered and
  modulated ordered phases.}
\label{fig:gpd}
\end{figure}

The concept of a Lifshitz point was introduced more than 25 years ago
\cite{HLS75b}. It is a point in the phase diagram at which a
disordered phase, a spatially homogeneous ordered phase, and a
spatially modular ordered phase meet. A typical phase diagram with a
Lifshitz point is depicted in Fig.~\ref{fig:gpd}. In the case of a
ferromagnet, the disordered and uniform ordered phases are the usual
paramagnetic and ferromagnetic phases. The order of the latter
corresponds to an infinite `modulation' wave-length, i.e., a
modulation wave-vector $\boldsymbol{q}_0=0$. The transition that
occurs upon crossing the phase boundary between the disordered and the
uniform ordered phases is \emph{continuous} (`second-order
transition').  Thus each point on the line $T=T_{\mathrm c}(g)$
separating these two phases and emerging from the Lifshitz point is a
critical point. The variable $g$ is a second (intensive) thermodynamic
variable besides temperature $T$. What it stands for depends on the
type of system considered: In the case of organic crystals like
TTF-TCNQ \cite{AD77,HZW80}, $g$ corresponds to pressure; in the cases
of the much studied magnet MnP \cite{SBOC81,ZSK00} and the so-called
ANNNI model \cite{FS}, which we will both briefly consider below, $g$
stands for the magnetic field component perpendicular to the order
parameter and a ratio of an antiferromagnetic to a ferromagnetic
interaction coefficient, respectively. The important point to remember
is that $g$ \emph{does not couple directly to the order parameter}.
(If it did, a small change $g\to g+\delta g$, $T\to T_{ \mathrm c}
+\delta T$ along any direction in the $gT$ plane---and specifically
along $T_{\mathrm c}(g)$--- would destroy the critical behavior; i.e.,
there would not be a critical line.) For this reason, $g$ is commonly
called a \emph{nonordering} thermodynamic field.

The term `thermodynamic field' is probably not too familiar among
high-energy physicist, and should by no means be confused with what is
meant by a field in field theory. Its usage was suggested in a seminal
paper \cite{GW70} by Griffiths and Wheeler for intensive thermodynamic
variables like $T$, magnetic field, pressure, and chemical potential
that take \emph{the same values on both sides of (in general,
  first-order) bulk phase transition}. Their thermodynamic
counterparts are the densities of extensive variables like the
magnetization, energy, and entropy densities that normally jump at the
transition (though not necessarily all of them) since their values in
the respective pure bulk phases differ.

In the modulated ordered phase, the order is characterized by a
nonzero modulation wave vector $\boldsymbol{q}_0$, which varies with
$T$ and $g$. The transition between the disordered and modulated
ordered phase is continuous as well. Hence the Lifshitz point divides
the critical line $T_{\mathrm c}(g)$ into two sections. The transition
between the homogeneous and modulated ordered phases can be of first
or second order; for models with a scalar order parameter it is
generically discontinuous, for specific models with vector order
parameters it is found to be continuous \cite{CL95}.

Let us also note that the modulation wave-vector $\boldsymbol{q}_0$
does not necessarily have to vanish at the Lifshitz point. To see
this, recall that in the case of antiferromagnets, the order parameter
which acquires a nonzero value in the uniform ordered phase is
\emph{not} the magnetization but the \emph{staggered magnetization}.
If we consider lattice models, this is given by the average of the sum
of all spins on one (up-spin) sublattice minus the sum of those on the
other (down-spin) sublattice. Thus the homogeneous ordered phase does
not correspond to a homogeneous magnetization density but to one
modulated with the corresponding nonzero wave-vector at the boundary
of the Brillouin zone. This value of the modulation wave-vector
$\boldsymbol{q}_0$ applies in particular to the Lifshitz point, so
$\boldsymbol{q}_{0,{\mathrm L}}\ne\boldsymbol{0}$. For the sake of
simplicity, we shall always use ferromagnetic language in the sequel,
taking $\boldsymbol{q}_{0,{\mathrm L}}$ to vanish.

\subsection{Critical exponents and continuum models}
\label{sec:criexp}

To reach the Lifshitz point, both the temperature $T$ and the
nonordering field $g$ must be fine-tuned. This tells us that $g$ must
correspond to a relevant variable in the RG sense. We denote the
associated crossover exponent as $\varphi$. It describes the behavior
of the critical line $T_{\mathrm c}(g)$ in the vicinity of L: As
indicated in Fig.~\ref{fig:gpd}, we have
\begin{equation}
  \label{eq:varphidef}
  \delta T_{\mathrm c}\equiv T_{c}(g)-T_{\mathrm L}\sim
|\delta g|^{1/\varphi}\;,\quad\delta g\equiv g-g_{\mathrm L}\;.
\end{equation}
In order to describe the behavior of the modulation wave-vector near
L, one introduces a wave-vector exponent $\beta_q$ via
\begin{equation}
  \label{eq:betaqdef}
  \delta q_0\equiv q_0-q_{0,{\mathrm L}}\sim |\delta g|^{\beta_q}\;.
\end{equation}
There are further critical exponents that are needed to characterize
the critical behavior at Lifshitz points. We will introduce some of
these below.

How a Lifshitz point can occur can be easily understood within Landau
theory. Landau theory leads one to consider the following natural
generalization of the usual $\phi^4$ model with the Hamiltonian
(Euclidean action)
\begin{equation}\label{eq:Ham}
{\mathcal{H}}={\int}\!{d^d}x
{\left\{
\frac{1}{2}\,{({{\nabla}_\perp}
 \bphi )}^2
+\frac{\rho_0}{2}\,{({{\nabla}_\parallel}
 \bphi )}^2
+\frac{\sigma_0}{2}\,{(\triangle_\parallel
 \bphi )}^2
+\frac{\tau_0}{2}\,
\bphi^2+\frac{u_0}{4!}\,|\bphi |^4\right\}}\;.
\end{equation}
Here $\bphi=(\phi^\alpha)$ is an $n$-component order-parameter field.
The $d$-dimensional position vector $\bx =(\bxp,\bxperp)$ has an
$m$-dimensional `parallel' component $\bxp$ and a $(d-m)$-dimensional
`perpendicular' one $\bxperp$. The coefficients of the squared
gradient terms, and those of the Hamiltonian's other interaction
terms, generally depend on the thermodynamic variables $T$ and $g$.
We have assumed that the squared gradient terms involve just two
distinct coefficients, where the one of $(\nabla_\perp\bphi)^2$
remains positive and has been transformed to unity by means of an
appropriate choice of the amplitude of $\bphi$. Its parallel
counterpart, $\rho_0$, is permitted to change sign. To ensure
stability when $\rho_0<0$, the term $(\triangle_\|\bphi)^2$ with a
positive coefficient $\sigma_0>0$ has been added. The critical line
between the disordered and homogeneously ordered phases is given in
Landau theory by $\tau_0=0$ with $\rho_0>0$. The Lifshitz point is
located at $\tau_0=\rho_0=0$ in this classical approximation.

Near the Lifshitz point the interaction coefficients can be expanded
about $T=T_{\mathrm L}$ and $g=g_{\mathrm L}$. For the coefficients that
remain positive, i.e.\ $u_0$ and $\sigma_0$, the expansions may be
truncated at zeroth order, so that $u_0$ and $\sigma_0$ become
independent of $T$ and $g$. The deviations $\delta\tau_0\equiv
\tau_0-\tau_{0,{\mathrm L}}$ and $\delta \sigma_0=\sigma_0-\sigma_{0,{\rm
    L}}$ change sign at L (where $\tau_{0,{\mathrm L}}$ and
$\sigma_{0,{\mathrm L}}$ vanish in Landau theory); their expansions must be
retained to linear order in $T-T_{\mathrm L}$ and $\delta g$.

\section{What is interesting about studying critical behavior at
  Lifshitz points}
\label{sec:motiv}

One reason for the ongoing interest in critical behavior at Lifshitz
points is the wealth of distinct physical systems with such
multi-critical points. They range from magnetic ones
\cite{SBOC81,ZSK00,FurRef}, ferroelectric crystals \cite{VS92}, and
charge-transfer salts \cite{AD77,HZW80} to liquid crystals
\cite{CL76}, systems undergoing structural phase transitions
\cite{AM80} or having domain-wall instabilities \cite{LN79}, and the
so-called ANNNI model \cite{FS}. Lifshitz points have been
discusssed recently even in the context of superconductors
\cite{Nog00} and polymer blends \cite{BMLSM95}.

From a more general perspective, the problem is interesting because it
provides well-defined clear examples of systems exhibiting
\emph{anisotropic scale invariance (ASI)}. At conventional critical
points, scaling operators ${\mathcal O}$ such as the order parameter
$\bphi$ and the energy density transform asymptotically as ${\mathcal
  O}(\ell \bx)=\ell^{-\Delta_{\mathcal O}}\,{\mathcal O}(\bx)$ under
scale transformations, with a (in general nontrivial) scaling
dimension $\Delta_{\mathcal O}$. In the case of ASI, the position
coordinates $\bx$, or the position and time coordinates (in the case
of time-dependent phenomena), divide into two (or more) groups, say
$\bx=(\bxp,\bxperp)$, that must be scaled with different powers of the
scale factor $\ell$ to recover ${\mathcal O}(\bx)$; one has
\begin{equation}
  \label{eq:asi}
  {\mathcal O}(\ell^\theta \bxp,\ell \bxperp)=
 \ell^{-\Delta_{\mathcal O}}\,{\mathcal O}(\bx)\;,
\end{equation}
with an anisotropy exponent $\theta$ different from unity. That ASI
applies at Lifshitz points is evident from the Hamiltonian
(\ref{eq:Ham}): The momentum-space two-point vertex function of the
free theory obviously behaves at the Lifshitz point $\tau_0=\rho_0=0$
as
\begin{equation}
  \label{eq:Gamma2}
  \tilde{\Gamma}^{(2)}({\boldsymbol{q}}_\|={\boldsymbol{0}},
      {\boldsymbol{q}}_\perp\to{\boldsymbol{0}})
\sim q_\perp^{2-\eta_{{\mathrm L}2}}\quad\mbox{and}\quad 
  \tilde{\Gamma}^{(2)}({\boldsymbol{q}}_\|\to{\boldsymbol{0}},
{\boldsymbol{q}}_\perp={\boldsymbol{0}})\sim q_\|^{4-\eta_{{\mathrm L}4}}\;,
\end{equation}
where the analogs $\eta_{{\mathrm L}2}$ and $ \eta_{{\mathrm L}4}$ of the
usual correlation exponent $\eta$ take their mean-field value zero, but
differ from it beyond Landau theory. In terms of these, the anisotropy
exponent reads $\theta=({2-\eta_{{\mathrm L}2}})/({4-\eta_{{\mathrm L}4}})$.
The asymptotic validity of ASI means that two distinct correlation
lengths $\xi_\perp$ and $\xi_\|$ are needed to characterize the region
within which $\phi(\bx )$ is correlated to $\phi(\boldsymbol{0})$;
these diverge at $g=g_{\mathrm L}$ as functions of $\delta T=(T-T_{\rm
  L})/T_{\mathrm L}$ like $|\delta T|^{-\nu_{{\mathrm L}2}}$ and $|\delta
T|^{-\nu_{{\mathrm L}4}}$ with different exponents $\nu_{{\mathrm L}2}$ and
$\nu_{{\mathrm L}4}=\theta\,\nu_{{\mathrm L}2}$, respectively.

A familiar arena of ASI are \emph{dynamic critical phenomena near
  equilibrium} \cite{HH77,Folk}. In their case the time $t$ and
position $\bx$ must be scaled differently; the analog of
Eq.~(\ref{eq:asi}) becomes $\bphi(\ell
\bx,\ell^{\mathfrak{z}}t)=\ell^{-\Delta_\phi}\,\bphi(\bx,t)$ for the
order parameter, and the role of $\theta$ is played by the dynamic
exponent $\mathfrak{z}$. Their simplifying feature is that detailed
balance and fluctuation-dissipation theorems hold. This entails that
the stationary states of the dynamics are guaranteed to be thermal
equilibrium ones described by an a priori known Hamiltonian. Hence the
problem of the steady-state correlations splits off from the dynamics;
all critical exponents of static origin can be determined without
dealing with dynamics, only genuine dynamic properties require the
analysis of the dynamic field theory.

ASI is abundant in \emph{non-equilibrium} systems such as driven
diffusive systems \cite{SZ95} and surface growth processes
\cite{Kru97}, which have attracted considerable attention during the
past decade. In their case detailed balance is not normally
valid. Finding the steady-state solutions therefore is a nontrivial
task, requiring the investigation of the long-time limit of the
dynamics. This must be solved before the ASI of the steady-state
correlations can be investigated.

Two features make critical behavior at $m$-axial Lifshitz points a
very attractive stage for the study of ASI: (i) One can stay entirely
within the realm of equilibrium statistical physics, and (ii) the
class of models (\ref{eq:Ham}) involves a parameter, $m$, that can be
varied. Some time ago Henkel \cite{Hen97,Hen99} argued that ASI should
imply additional invariances, just as scale and rotational invariance
in local field theories lead to the larger symmetry group of conformal
transformations \cite{Pol70,BPZ84,Car87}. He made predictions for the
form of the scaling function of the pair correlation function at
criticality, which are in conformity with analytic results for certain
spherical models \cite{FH93}. Recent Monte Carlo results for an
equilibrium \cite{PH01} and non-equilibrium systems \cite{HPGL01}
appear to support these claims. Yet the general validity of such
additional invariances has not been shown, nor is there a good
understanding of their origin or the conditions under which they hold.
For example, in Ref.~\cite{Hen97} the assumption that the anisotropy
exponent takes the rational values $\theta=2/\wp$, $\wp\in \mathbb{N}$, is
needed to ensure that the considered sub-algebra closes. However, in
the case of the uniaxial Lifshitz point in $d=3$ dimensions studied by
means of simulations \cite{PH01}, this condition is unlikely  to be fulfilled
since the $\epsilon$ expansion \cite{DS00a,SD01} reveals that $\theta$
differs from $1/2$ at order $\epsilon^2$, although the $d=3$ estimate
$\theta\simeq 0.487$ it yields for $m=n=1$ is pretty close to it.
Building on the field-theoretic analysis described in
Refs.~\cite{DS00a} and \cite{SD01}, and outlined below, one should be
able to clarify whether the suggested `generalized conformal
invariance' and the predicted form of two-point scaling functions hold
indeed.

Before turning to the RG analysis of the models (\ref{eq:Ham}), let us
discuss two examples of systems with a Lifshitz point.

\section{Two examples of systems with Lifshitz points: the ANNNI model
  and MnP}
\label{sec:examp}

\begin{figure}[htbp]
\begin{center}
\includegraphics[width=60mm]{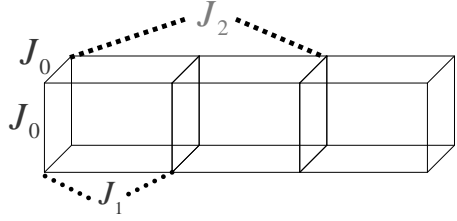}
\end{center}
\caption{ANNNI model. Along one axis the spins are coupled via
  ferromagnetic nearest-neighbor bonds of strength $J_1$ and
  antiferromagnetic axial next-nearest neighbor bonds of strength
  $|J_2|\equiv -J_2$. Along the other directions there are only
  ferromagnetic nearest-neighbor interactions of strength $J_0$.}
\label{fig:annni}
\end{figure}

We begin with the axial-next-nearest-neighbor Ising (ANNNI) model,
defined through the Hamiltonian
\begin{equation}
  \label{eq:Hanni}
  {\mathcal H}_{\mathrm{ANNNI}}=
  -\frac{J_0}{k_{\mathrm B}T}\sum_{i,\delta_\perp}s_i\,s_{i+\delta_\perp}
  -\frac{J_1}{k_{\mathrm B}T}\sum_{i,\delta_\|}s_i\,s_{i+\delta_\|}
  -\frac{J_2}{k_{\mathrm B}T}\sum_{i,\delta_\|'}s_i\,s_{i+\delta_\|'}\;,
\end{equation}
where $s_i\pm 1$ are Ising spins residing on the sites $i$ of the
cubic lattice illustrated in Fig.~\ref{fig:annni}. Here $\delta_\|$
and $\delta_\|'$ are nearest-neighbor (nn) and next-nearest-neighbor
(nnn) displacements along one axis, while $\delta_\perp$ denote nn
displacements along any of the other (perpendicular) directions.
The nn couplings are ferromagnetic, $J_0>0$, $J_1>0$; the nnn bond is
antiferromagnetic, $J_2<0$.

\begin{figure}[htbp]
\begin{center}
\includegraphics[width=8cm]{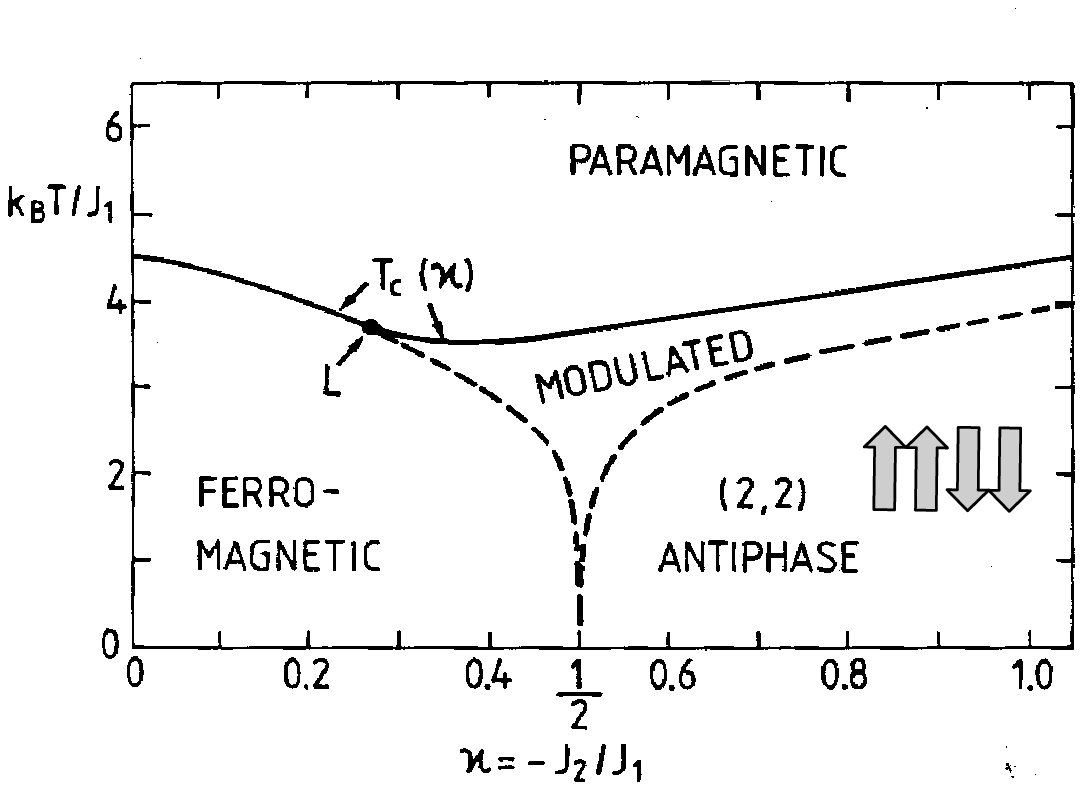}
\end{center}
\caption{Phase diagram of the three-dimensional  ANNNI model, taken
  from Ref.~\cite{Sel92} (with permission kindly granted by its
  author); to illustrate the ordering in the (2,2) phase the inset
  with the arrows has been added.  $\mathrm L$ is a uniaxial Lifshitz
  point.}
\label{fig:annnipd}
\end{figure}

The phase diagram of the three-dimensional ANNNI model is depicted in
Fig.~\ref{fig:annnipd}.
Obviously, the nonordering field $g$ here translates into the ratio
$\kappa=-J_2/J_1$. The phase labeled (2,2), also called $\langle
2\rangle$ structure, corresponds to an antiferromagnetic ordering of
the kind indicated by the arrows in the inset.  The region marked
`modulated' actually has considerably more structure: From the
multi-phase point $k_{\mathrm B}T/J_1=0$, $\kappa=1/2$, commensurate
phases of type $\langle 3\rangle$ and $\langle 2^p3\rangle$,
$p=1,2,\ldots,\infty$, split off. Here $\langle 3\rangle$  denotes a periodic layer sequence of 3-bands, while  $\langle
2^p3\rangle$ signifies a periodic layer sequence of $p$ 2-bands followed by a
3-band, where a $k$-band means that the magnetization has the same
sign in $k$ successive layers. Though interesting, these details
cannot be expounded here because they are off our main topic. They can
be found in the literature \cite{Sel92,FS}. The essential point for
us to note is that L is a \emph{uniaxial} ($m=1$) Lifshitz point. It has been
repeatedly studied via Monte Carlo simulations to determine the values
of the critical exponents for the case $m=n=1$, $d=3$
\cite{Sel78,Sel80,KS85,PH01}. 

\begin{figure}[htbp]
\begin{center}
\includegraphics[angle=-90,width=8cm]{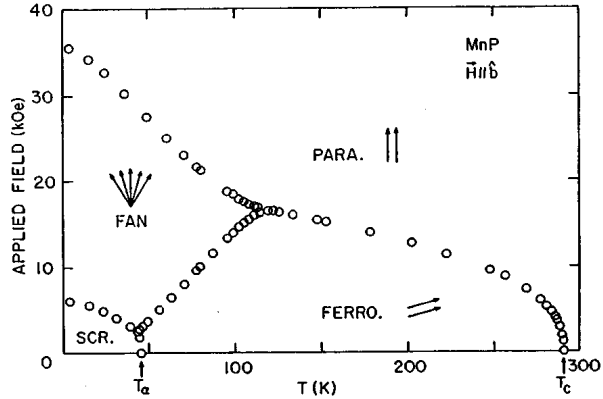}
\end{center}
\caption{Phase diagram of MnP for a magnetic field parallel to the
  $\boldsymbol{b}$ axis, according to Refs.~\cite{SBOC81} and
  \cite{ZSK00}.}
\label{fig:mnp}
\end{figure}

An experimentally much studied system is the orthorhombic metallic
compound MnP \cite{SBOC81,ZSK00,YCS84}. Its phase diagram differs,
depending on whether the magnetic field $\boldsymbol{H}$ is directed
along the $\boldsymbol{a}$, $\boldsymbol{b}$, or $\boldsymbol{c}$
crystal axis. Both for $\boldsymbol{H}\| \boldsymbol{a}$ and
$\boldsymbol{H}\| \boldsymbol{b}$ a uniaxial Lifshitz point is found,
but not for $\boldsymbol{H}\| \boldsymbol{c}$. The phase diagram of
Ref.~\cite{SBOC81} for the case $\boldsymbol{H}\| \boldsymbol{b}$ is
reproduced in Fig.~\ref{fig:mnp} (with permission kindly granted by
the first author, Y.\ Shapira ).
The magnetic field component $\boldsymbol{H}\cdot\boldsymbol{b}$ plays
the role of the thermodynamic field $g$. In the phase labeled `para'
the magnetic moments (`spins') $\boldsymbol{s}(\boldsymbol{x})$ are
aligned along the $\boldsymbol{H}\propto \boldsymbol{b}$ direction. In
the `ferro' phase the magnetization
$\boldsymbol{m}=\langle\boldsymbol{s}\rangle$ has a component along
the $\boldsymbol{c}$ axis and hence is tilted with respect to
$\boldsymbol{b}$. The component $\boldsymbol{c}\cdot\boldsymbol{s}$
plays the role of the order parameter; since $\boldsymbol{a}$ is a
very hard axis, $\boldsymbol{s}\cdot\boldsymbol{a}$ can be ignored
\cite{YCS84}. In the fan phase the spins rotate in the $bc$ plane 
(but make no full turn as in the screw phase
`SCR' not considered here); there is modulated order with a modulation
wave-vector ${\boldsymbol{q}}_0 \propto\boldsymbol{a}$. The ferro-fan
transition is first order. The meeting point of the para, ferro, and
fan phases is an $m=n=1$ Lifshitz point.

\section{Dimensionality expansions and renormalization group analysis}
\label{sec:RGA}

Setting $\rho_0=\tau_0=0$ in the Hamiltonian (\ref{eq:Ham}), we see
that $q_\perp$ scales as $q_\|^2$ at the Lifshitz point of the
Gaussian ($u_0{=}0$) theory. By dimensional analysis we have
$[x_\perp]=\mu^{-1}$, $[x_\|]=\sigma_0^{1/4}\mu^{-1/2}$,
$[\tau_0]=\mu^2$, $[\rho_0]=\sigma_0^{1/2}\mu$, and
$[u_0]=\sigma_0^{m/4}\mu^{\epsilon}$ with $\epsilon\equiv d^*(m)-d$, where
\begin{equation}
  \label{eq:dstar}
   d^*(m)=4+\frac{m}{2}\;,\quad m\le 8\;.
\end{equation}
From the given engineering dimensions we can read off how the
interaction coefficients transform under the scale transformation
$\mu\to\mu\ell$. The coupling constant $u_0$ becomes marginal at
$d=d^*(m)$, the upper critical dimension (UCD). Note that $m=8$
implies $d^*=m=8$; i.e., this is the case of the isotropic Lifshitz
point in which only the parallel components of position and momentum
remain.

\begin{figure}[htb]
\begin{center}
\includegraphics[width=100mm]{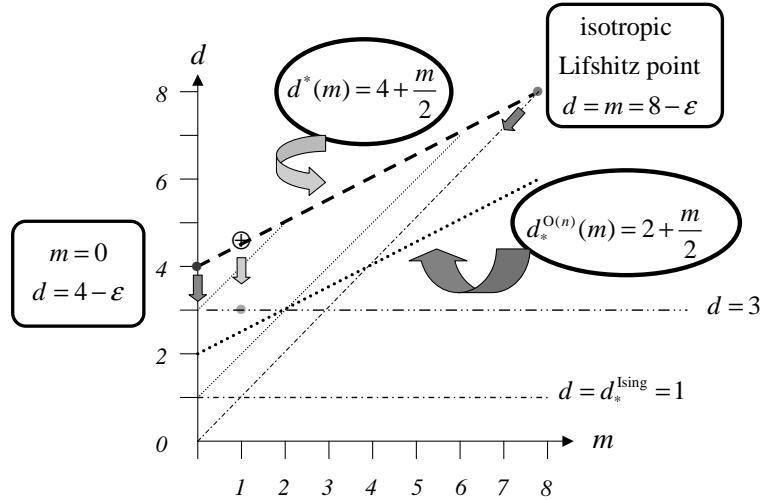}
\end{center}
\caption{$md$ plane with the line of upper critical dimensions
  $d^*(m)$. For further explanations see main text.}
\label{fig:de}
\end{figure}

One can analytically continue the parallel and perpendicular momentum
integrals in their respective dimensions $m$ and $d-m$, and hence take
those as continuously varying. Ideally one would like to determine the
expansions in $m$ and $d-m$ of the critical exponents and other
universal quantities about any point of the line $d^*(m)$.  The
situation is illustrated in Fig.~\ref{fig:de}.
The point $(m,d)=(0,4)$ marks the UCD of the standard (isotropic)
$|\bphi|^4$ model; analyzing it via the conventional $\epsilon=4-d$
expansion means to move away from this point along the path indicated
by $\Downarrow$. More generally, $\epsilon$ expansions at fixed $m$
correspond to paths parallel to the $d$-axis. How to reach the
physically interesting point $(1,3)$ along such a path is indicated.
In the case of the isotropic $(m{=}d)$ Lifshitz point, things are
different: The conventional expansion in $\epsilon_8\equiv 8-d$ is
along the diagonal $m=d$ \cite{DS02,HLS75b}.

We must also keep in mind an obvious condition points $(m,d)$ to which
extrapolations make sense must satisfy: $d$ must be larger than
$d_*(m,n)$, the lower critical dimension (LCD) below which a
Lifshitz---or, if $m=0$, a critical---point cannot occur. Since the
LCD of the Ising model with short-range interactions is $d_*(0,1)=1$,
we clearly must have $d>1$. In the $n>1$ case with the continuous $O(n)$ symmetry,
familiar spin-wave arguments and an analysis based on an adequate
nonlinear sigma model \cite{GS78} show that the ordered phase becomes
thermally unstable at any $T>0$ if $d\le d_*^{O(n)}(m)\equiv
2+\frac{m}{2}$ (cf.\ Fig.~\ref{fig:de}).

The existence of a Lifshitz point with $0<m<8$ also requires that the
modulated ordered phase be thermally stable. In the critical region of
this phase, fluctuations of the Fourier components
$\boldsymbol{\phi}_{\boldsymbol{q}}$ of the order parameter with
wave-vectors $\boldsymbol{q}\simeq \pm {\boldsymbol{q}}_0$ are
dominant. As pointed out in Ref.~\cite{GP76}, one therefore expects
that an $n$-component system with a helical structure behaves
critically as a $2n$-component $\phi^4$ model whose $\phi^4$ terms are
$O(n)$ but not $O(2n)$ symmetric. Specifically for
$n=1$, one arrives at an anisotropic two-component model. It has
(besides others) an $O(2)$ symmetric fixed point that is believed to
be stable \cite{GP76}. This suggests that the long-range order (LRO)
of the modulated phase should be destroyed in $d\le 2$ dimensions by
thermal fluctuations at any temperature $T>0$. Furthermore, if the
presumed isotropy of the Hamiltonian (\ref{eq:Ham}) in the parallel subspace
can indeed be taken for granted, one can exploit the invariance under
arbitrary rotations along the lines of Mermin and Wagner \cite{MW66}
to show the absence of a helical phase with orientational LRO for
$m\le d\le m+1$ \cite{Lub72,ML78}.%
\footnote{In Ref.~\cite{ML78} the fact that the one-loop shift of
  $T_{\mathrm{c}}(g)$ on the helicoidal section of the critical line
  diverges in the infrared if $m\le d\le m+1$ is interpreted as
  signaling the absence of helical LRO. However, this property alone
  is not sufficient to rule out LRO: The corresponding shift of
  $T_{\mathrm{c}}$ of the standard one-component $\phi^4$ model is
  infrared divergent for $d\le 2$. Nevertheless, the $d=2$ Ising model
  has a ferromagnetic low-temperature phase.}

The goal of expanding in $m$ and $d$ about a general point on the line
$d^*(m)$ was envisaged already in Ref.~\cite{HLS75b}. Yet its
realization turned out to be extremely difficult. Computing the
$\epsilon$ expansions of the critical exponents for general values of
$m$ to first order in $\epsilon$ is easy. Their $\Or(\epsilon)$
coefficients are independent of $m$, and essentially determined by
combinatorial factors. (The operator product expansion is structurally
similar to the one that applies to the standard $|\phi|^4$ theory.
Upon generalizing Cardy's analysis in Ref.~\cite{Car96b} to the $m>0$
case, one can determine these coefficients without having to work out
the Feynman diagrams in detail.) The real challenges start at order
$\epsilon^2$.

The first results to order $\epsilon^2$ for general $m$ were given in
1977 by Mukamel \cite{Muk77}. The $\epsilon^2$ term of $\eta_{{\mathrm
    L}2}$ he determined by means of Wilson's momentum-shell
integration method was independent of $m$, that of $\eta_{{\mathrm
    L}4}$ differed from it by a simple $m$-dependent factor. He also
computed $\beta_q$ to $\Or(\epsilon^2)$ for $m<6$. For $m=1$, these
results were confirmed by Hornreich and Bruce \cite{HB78}. Utilizing
also Wilson's technique, Sak and Grest \cite{SG78} did an independent
calculation; because of the severe technical difficulties they
encountered for general $m$ they confined themselves to $m=2$ and
$m=6$. Their results are at variance with Mukamel's; in particular,
the $\epsilon^2$ coefficients of $\eta_{{\mathrm L}2}(m,n)$ for $m=2$
and $m=6$ are not equal, and hence \emph{not} $m$-independent.

The application of modern field-theory RG approaches to the problem
began in 1998.  Mergulh{\~a}o and Carneiro formulated normalization
conditions and derived RG equations for the renormalized theory
\cite{MC98}. In a subsequent paper \cite{MC99}, they reproduced Sak
and Grest's results for $\eta_{{\mathrm L}2}$ and $\eta_{{\mathrm
    L}4}$ with $m=2$ and $m=6$, and performed a two-loop calculation
for these two values of $m$. They fixed the perpendicular dimension
$d-m$ at $d^*(2)-2=3$ and $d^*(6)-6=1$, taking the parallel one as
$2-\epsilon_\|$ and $6-\epsilon_\|$, respectively. The respective
paths starting from the points $(m,d)=(2,5)$ and $(6,7)$ are indicated
in Fig.~\ref{fig:de}.

In two recent papers \cite{DS00a,SD01} Shpot and myself have been able
to perform a full two-loop calculation for general $m\in (0,8)$ and to
determine the $\epsilon$ expansions of all critical, crossover,
wave-vector, and correction-to-scaling (Wegner) exponents to
$\Or(\epsilon^2)$. The results are analytical except that the two-loop
terms of the required renormalization functions and the series
expansion coefficients of the exponents' $\epsilon^2$ terms involve
four well-defined single integrals $j_\phi(m)$, $j_\sigma(m)$,
$j_\rho(m)$, and $J_u(m)$ which for general $m$ we have not been able
to evaluate analytically, though for the special values $m=0$, 2, 6,
and 8. For other values of $m$ theses integrals can be computed
numerically, as we did for $m=1$, 2, \ldots, 7.

These results stand a number of nontrivial checks. First of all, they
reduce to the well-known results for the standard $|\bphi|^4$ model
in the limit $m\to 0$.%
\footnote{More precisely, this applies to quantities retaining their
  physically significance for $m=0$, like $\eta_{{\mathrm
      L}2}(m{=}0)\equiv \eta$ and $\nu_{{\mathrm L}2}(m{=}0)\equiv
  \nu$. The series expansion coefficients of other exponents that are
  not required (nor have a known immediate physical meaning) in the
  $m=0$ theory, such as $\eta_{{\mathrm L}4}(m)$, may nevertheless
  have finite $m\to 0$ limits \cite{SD01,DS02}.}
Second, the analytical results they yield for $m=2$ and $m=6$ are
consistent with and extend those of Sak and Grest \cite{SG78} and of
Mergulh{\~a}o and Carneiro \cite{MC99}. [The original
$\Or(\epsilon^2)$ results of Ref.~\cite{MC99} for $\nu_{{\mathrm
    L}2}(m{=}2,6)$ and $\nu_{{\mathrm L}4}(m{=}2,6)$ disagreed with
ours but become identical to those upon elimination of two minor
computational errors.] The isotropic case $m=d$ provides a third,
highly \emph{nontrivial}, check. To see this, note that the $\epsilon$
expansions of the critical exponents $\lambda=\eta_{{\mathrm L}4}$,
\ldots is of the form
\begin{equation}
\label{eq:lambda}
  \left.\lambda(n,m,d)\right|_{d=d^*(m)-\epsilon}
                =\lambda_0+\lambda_1(n)\,\epsilon
                +\lambda_2(n,m)\,\epsilon^2+\Or(\epsilon^3)\;,
\end{equation}
i.e., the $m$-dependence starts at order $\epsilon^2$. To compare with
the $\epsilon_8\equiv 8-d$ expansion for the isotropic case, we set
$m=d=8-\epsilon_8$ (which gives $\epsilon=\epsilon_8/2$) in
Eq.~(\ref{eq:lambda}). Hence the expansions of the critical exponents
$\lambda$ of the isotropic Lifshitz point to quadratic order in
$\epsilon_8$ should result from the right-hand side via the
replacements $\epsilon\to\epsilon_8/2$ and
$\lambda_2(n,m)\to\lambda_2(n,8-)$. The so-obtained $\epsilon_8$
expansions of the exponents $\lambda=\eta_{{\mathrm L}4}(n,d,d)$,
$\nu_{{\mathrm L}4}(n,d,d)$, $\varphi(n,d,d)$, $\beta_q(n,d,d)$, and
the correction-to-scaling exponent $\omega_{{\mathrm L}4}(n,d,d)$ have
been verified by means of a direct field-theo\-retic investigation of
the $m=d$ case \cite{DS02}. Yet, there is one group of authors
\cite{dAL} who obtained---and still favor---results at variance with
ours in Refs.~\cite{DS00a,SD01}; their findings and criticism have
been refuted in Refs.~\cite{DS01a,DS02}.

The origin of the technical difficulties which had prevented two-loop
calculations for so long and caused the mentioned long-standing
controversies can be traced back to the difficult form of the free
propagator $G(\bx)$ at the Lifshitz point $\tau_0=\rho_0=0$. In the
isotropic cases $m=0$ and $m=d$, $G(\bx)$ is a simple power of $x$.
However, for general $m$ (and $\sigma_0=1$) it becomes
\begin{equation}
  \label{eq:fprop}
  G(\bxp,\bxperp )=\int_{\boldsymbol{q}}
   \frac{\e^{i\boldsymbol{q}\cdot\bx}}{q_\|^4+q_\perp^2}
 =x_\perp^{-2+\epsilon}\,\Phi_{m,d}{\big(x_\|\,x_\perp^{-1/2}\big)}\;,
\end{equation}
where $\Phi_{m,d}$ are extremely complicated scaling functions of the
form
\begin{equation}
  \label{eq:Phimd}
  \Phi_{m,d}(\upsilon)=C^{(1)}_{m,\epsilon}\,{{_{1\!}{\mathrm
        F}_{2}}}\!\left( 1{-}\frac{\epsilon}{ 2};\frac{1}{2},\frac{2{+}m}{
      4};\frac{\upsilon^4}{ 64}\right)
-C^{(2)}_{m,\epsilon}\, {\upsilon^2}\,
{_{1\!}{\mathrm F}_{2}}\!\left(\frac{3}{2}{-}\frac{\epsilon}{ 2};
\frac{3}{ 2},1{+}\frac{m}{ 4};\frac{\upsilon^4}{ 64}\right).
\end{equation}
The latter are generalizations of generalized hypergeometric functions
known as Fox-Wright $\,_{1\!}\psi_{1}$ functions \cite{DS00a,SD01}.
They have asymptotic expansions in powers of $\upsilon^{-4}$. On the
line $d^*(m)$ they can be expressed in terms of modified Struve and
Bessel functions. The cases $m=2$ and $m=6$ are special in that
remarkable simplifications occur: The asymptotic expansions of
$\Phi_{2,5}$ and $\Phi_{6,7}$ truncate at zeroth or first order in
$\upsilon^{-4}$, respectively, and these functions reduce to a simple
exponential and a similar elementary function, respectively. This is
what makes the calculation analytically feasible for $m=2,6$.

To set up the RG for general $m$ and $d=d^*(m)-\epsilon$, one can
introduce a renormalized field $\bphi_{\mathrm{ren}}$ and renormalized
coefficients $\sigma$, $\tau$, $\rho$, and $u$ via
\begin{eqnarray}
  \label{eq:rep} 
&& \bphi=Z_\phi^{1/2}\bphi_{\mathrm{ren}}\,,\quad 
   \sigma_0=Z_\sigma\,\sigma\,,\quad 
   \delta\tau_0=\mu^2 Z_\tau \tau\,, \nonumber\\&&
   \delta\rho_0\,\sigma_0^{-1/2}=\mu
Z_\rho\rho\,,\quad  u_0\,\sigma_0^{-m/4}=\mu^\epsilon Z_u u\,,
\end{eqnarray}
and determine the renormalization factors by minimal subtraction of
poles.  The residues of the $Z$s at two-loop order can be either
analytically calculated or expressed in terms of single integrals
whose integrands involve scaling functions such as $\Phi_{m,d^*(m)}$.

Let us illustrate this by means of the example of the two-loop graph
\raisebox{-0.25em}{\includegraphics[width=10mm]{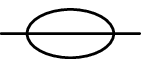}}, which is
$\propto G(\bx)^3$ in position space. To determine the Laurent
expansion of this distribution, we apply it to a test function
$f(\bx_\|,\bx_\perp)$, substituting Eq.~(\ref{eq:fprop}) for $G(\bx )$
and making the variable transformation $\bx_\|\to
\boldsymbol{\upsilon}=\bx_\|/\sqrt{x_\perp}$. One is led to an
integral of the form
$\int_0^{\infty}\!dx_\perp\,x_\perp^{2\epsilon-3}\,g_{\epsilon,m}(x_\perp)=
(4\epsilon)^{-1}g_{0,m}''(0)+\Or(\epsilon^0)$. Upon evaluating this
one sees that $G(\bx )^3$ has poles $\propto
\triangle_\|^2\,\delta(\bx)/\epsilon$ and $\propto
\triangle_\perp\,\delta(\bx)/\epsilon$ whose residues are proportional
to
\begin{equation}
\label{eq:js}
 \int_0^\infty\!d\upsilon\,\upsilon^{m+3}\,\Phi_{m,d^*}(\upsilon)^3
  \quad\mbox{and}\quad
 \int_0^\infty\!d\upsilon\,\upsilon^{m-1}\,\Phi_{m,d^*}(\upsilon)^3\;,
\end{equation}
respectively. Up to a simple factor given in Eq.~(46) of
Ref.~\cite{SD01}, these are the integrals $j_\sigma(m)$ and
$j_\phi(m)$. The other integrals $j_\sigma(m)$ and $J_u(m)$ are of a
similar nature but involve besides $\Phi_{m,d^*}$ further scaling
functions of the free theory like the one pertaining to the
convolution $(\nabla_\|G*\nabla_\|G)(\bx)$ [$=$ line with an insertion
of $(\nabla_\|\bphi)^2$].

The RG equations implied by the reparametrizations (\ref{eq:rep}) and
the $\mu$-invariance of the bare theory can be exploited in a familiar
manner to derive the scaling properties of correlation and vertex
functions. For details the reader is referred to the original papers
\cite{DS00a,SD01}, where the $\epsilon$ expansions of the exponents are
also utilized to obtain estimates for the numerical values of the
critical exponents in the physical interesting cases with $d=3$. The
resulting values agree, in particular, remarkably well with the
results of a very recent Monte Carlo investigation \cite{PH01} of the
ANNNI model, but also with some (though not all) experimental results
and other theoretical estimates.

So far we assumed perfect isotropy of the derivative terms of the
Hamiltonian (\ref{eq:Ham}) in $\bx_\|$-space. This is unrealistic in
the case of lattice systems. Unless there is only one parallel
direction ($m=1$) or none, we should allow for a more general $q_\|^4$
term, making the replacement
\begin{equation}
\label{eq:aniso}
(\triangle_\|\bphi)^2\to w_a\,T^a_{ijkl}\,(\partial_i\partial_j\bphi)
  \partial_k\partial_l\bphi=(\triangle_\|\bphi)^2 
+w\,\sum_{i=1}^m(\partial_i^2\bphi)^2+\ldots
\end{equation}
in $\mathcal {H}$, where $T^a_{ijkl}$ are the respective fourth-rank
tensors compatible with the symmetry of the system. In the case of
cubic symmetry, the only other term on the right-hand side besides the
symmetric (first) contribution is the second one. To order
$\epsilon^2$, the dimensionless variable $w$ is found to be relevant
at the isotropic $w=0$ fixed point (the coefficient of the
$\epsilon^2$ term of its RG eigenexponent is positive) \cite{DSZ}.
Unfortunately, the RG analysis for general $w\ne 0$ that is required
to decide whether a new fixed point exists to which the flow leads is
rather complicated and still remains to be completed.

\section{Concluding remarks}
\label{sec:concl}

Almost 25 years after the discovery of Lifshitz points and the
introduction of the continuum models (\ref{eq:Ham}), systematic
field-theoretic RG analyses beyond one-loop order of the universality
classes these models represent have finally become feasible. This
could open the way to accurate quantitative field-theory
investigations and detailed comparisons with experimental results.
Hopefully, this will also trigger further experimental work on the
critical behavior at Lifshitz points. On the theoretical side, the
progress reported here in applying the field-theoretic RG to the study
of continuum models like (\ref{eq:Ham}) could lead to further insights
regarding the question as to whether and under what conditions
anisotropic scale invariance implies additional symmetries.

\begin{ack}
  My own work on Lifshitz points has benefited from the very pleasant
  and fruitful collaboration with Mykola Shpot. I am also indebted to
  Royce K.~P.\ Zia for joining us in our work \cite{DSZ} on the
  effects of anisotropic $q_\|^4$ terms.
  
  Finally, I would like to acknowledge the support by the Deutsche
  Forschungsgemeinschaft via Sonderforschungsbereich 237.
\end{ack}



\begin{thebibliography}{10}

\bibitem{Hor80}
R.~M. Hornreich:
\newblock \emph{J. Magn. Magn. Mater.} \textbf{15--18}  (1980) 387

\bibitem{Sel92}
W.~Selke:
\newblock In: \emph{Phase Transitions and Critical Phenomena}, eds. C.~Domb,
  J.~L. Lebowitz, vol.~15, pp. 1--72. Academic, London (1992)

\bibitem{MC98}
C.~Mergulh{\~a}o, Jr., C.~E.~I. Carneiro:
\newblock \emph{Phys. Rev. B} \textbf{58}  (1998) 6047

\bibitem{MC99}
C.~Mergulh{\~a}o, Jr., C.~E.~I. Carneiro:
\newblock \emph{Phys. Rev. B} \textbf{59}  (1999) 13 954

\bibitem{DS00a}
H.~W. Diehl, M.~Shpot:
\newblock \emph{Phys.\ Rev.\ B} \textbf{62}  (2000) 12 338;
\newblock cond-mat/0006007

\bibitem{SD01}
M.~Shpot, H.~W. Diehl:
\newblock \emph{Nucl. Phys. B} \textbf{612}  (2001) 340;
\newblock cond-mat/0106105

\bibitem{DS01a}
H.~W. Diehl, M.~Shpot:
\newblock \emph{J.\ Phys.\ A} \textbf{34}  (2001) 9101;
\newblock cond-mat/0106105

\bibitem{DS02}
H.~W. Diehl, M.~Shpot:
\newblock \emph{J.\ Phys.\ A} \textbf{??}  (2002) ??;
\newblock cond-mat/0204267

\bibitem{HLS75b}
R.~M. Hornreich, M.~Luban, S.~Shtrikman:
\newblock \emph{Phys. Rev. Lett.} \textbf{35}  (1975) 1678

\bibitem{AD77}
E.~Abraham, I.~E. Dzyaloshinskii:
\newblock \emph{Solid State Commun.} \textbf{23}  (1977) 883

\bibitem{HZW80}
C.~Hartzstein, V.~Zevin, M.~Weger:
\newblock \emph{J. Phys. (France)} \textbf{41}  (1980) 677

\bibitem{SBOC81}
Y.~Shapira, C.~Becerra, N.~Oliveira, Jr., T.~Chang:
\newblock \emph{Phys. Rev. B} \textbf{24}  (1981) 2780

\bibitem{ZSK00}
A.~Zieba, M.~Slota, M.~Kucharczyk:
\newblock \emph{Phys. Rev. B} \textbf{61}  (2000) 3435

\bibitem{FS}
M.~E. Fisher, W.~Selke:
\newblock \emph{Phys. Rev. Lett.} \textbf{44}  (1980) 1502;
\newblock \emph{Philos. Trans. R. Soc. London, Ser. A} \textbf{302}  (1981) 1

\bibitem{GW70}
R.~B. Griffiths, J.~C. Wheeler:
\newblock \emph{Phys. Rev. A} \textbf{2}  (1970) 1047

\bibitem{CL95}
P.~M. Chaikin, T.~C. Lubensky:
\newblock \emph{Principles of condensed matter theory}:
\newblock Cambridge University Press, Cambridge (GB) (1995)

\bibitem{FurRef}
 For a more complete set of references, see Refs.\ \cite{Hor80} and
  \cite{Sel92}.

\bibitem{VS92}
Y.~M. Vysochanski\u{\i}, V.~Y. Slivka:
\newblock \emph{Usp. Fiz. Nauk} \textbf{162}  (1992) 139:
\newblock [{S}ov. Phys. Usp. {\bf 35}, 123--134]

\bibitem{CL76}
J.~H. Chen, T.~C. Lubensky:
\newblock \emph{Phys. Rev. A} \textbf{14}  (1976) 1202

\bibitem{AM80}
A.~Aharony, D.~Mukamel:
\newblock \emph{J. Phys. C} \textbf{13}  (1980) L255

\bibitem{LN79}
J.~Lajzerowicz, J.~J. Niez:
\newblock \emph{J. Phys. (Paris) Lett.} \textbf{40}  (1979) L165

\bibitem{Nog00}
F.~S. Nogueira:
\newblock \emph{Phys. Rev. B} \textbf{62}  (2000) 14 559

\bibitem{BMLSM95}
F.~S. Bates, W.~Maurer, T.~P. Lodge, M.~F. Schulz, M.~W. Matsen:
\newblock \emph{Phys. Rev. Lett.} \textbf{75}  (1995) 4429

\bibitem{HH77}
B.~I. Halperin, P.~C. Hohenberg:
\newblock \emph{Rev. Mod. Phys.} \textbf{49}  (1977) 435

\bibitem{Folk}
 See R.\ Folk's contribution to these proceedings.

\bibitem{SZ95}
B.~Schmittmann, R.~K.~P. Zia:
\newblock \emph{Statistical Mechanics of Driven Diffusive Systems}, vol.~17 of
  \emph{Phase Transitions and Critical Phenomena}:
\newblock Academic, London (1995)

\bibitem{Kru97}
J.~Krug:
\newblock \emph{Adv. Phys.} \textbf{46}  (1997) 139

\bibitem{Hen97}
M.~Henkel:
\newblock \emph{Phys. Rev. Lett.} \textbf{78}  (1997) 1940

\bibitem{Hen99}
M.~Henkel:
\newblock \emph{Conformal invariance and critical phenomena}:
\newblock Texts and monographs in physics. Springer, Berlin (1999)

\bibitem{Pol70}
A.~M. Polyakov:
\newblock \emph{Pisma ZhETP} \textbf{12}  (1970) 538;
\newblock [JETP Lett.\ {\bf 12}, 381--383 (1970)]

\bibitem{BPZ84}
A.~A. Belavin, A.~M. Polyakov, A.~B. Zamalodchikov:
\newblock \emph{Nucl.\ Phys.\ B} \textbf{241}  (1984) 333

\bibitem{Car87}
J.~L. Cardy:
\newblock In: \emph{Phase Transitions and Critical Phenomena}, eds. C.~Domb,
  J.~L. Lebowitz, vol.~11, pp. 55--126. Academic, London (1987)

\bibitem{FH93}
L.~Frachebourg, M.~Henkel:
\newblock \emph{Physica A} \textbf{195}  (1993) 577

\bibitem{PH01}
M.~Pleimling, M.~Henkel:
\newblock \emph{Phys. Rev. Lett.} \textbf{87}  (2001) 125702

\bibitem{HPGL01}
M.~Henkel, M.~Pleimling, C.~Godr{\`e}che, J.-M. Luck:
\newblock \emph{Phys. Rev. Lett.} \textbf{87}  (2001) 265701

\bibitem{Sel78}
W.~Selke:
\newblock \emph{Z. Phys.} \textbf{29}  (1978) 133

\bibitem{Sel80}
W.~Selke:
\newblock \emph{J.\ Phys.\ C} \textbf{13}  (1980) L261

\bibitem{KS85}
K.~Kaski, W.~Selke:
\newblock \emph{Phys. Rev. B} \textbf{31}  (1985) 3128

\bibitem{YCS84}
C.~S.~O. Yokoi, M.~D. Coutinho-Filho, S.~R. Salinas:
\newblock \emph{Phys. Rev. B} \textbf{29}  (1964) 6341

\bibitem{GS78}
G.~S. Grest, J.~Sak:
\newblock \emph{Phys. Rev. B} \textbf{17}  (1978) 3607

\bibitem{GP76}
T.~Garel, P.~Pfeuty:
\newblock \emph{J. Phys. C} \textbf{9}  (1976) L246

\bibitem{MW66}
N.~D. Mermin, H.~Wagner:
\newblock \emph{Phys. Rev. Lett.} \textbf{17}  (1966) 1133

\bibitem{Lub72}
T.~C. Lubensky:
\newblock \emph{Phys.\ Rev.\ Lett.} \textbf{29}  (1972) 3885

\bibitem{ML78}
D.~Mukamel, M.~Luban:
\newblock \emph{Phys. Rev. B} \textbf{18}  (1978) 3631

\bibitem{Car96b}
J.~Cardy:
\newblock \emph{Scaling and Renormalization in Statistical Physics}:
\newblock Cambridge Lecture Notes in Physics. Cambridge University Press,
  Cambridge, Great Britain (1996)

\bibitem{Muk77}
D.~Mukamel:
\newblock \emph{J. Phys. A} \textbf{10}  (1977) L249

\bibitem{HB78}
R.~M. Hornreich, A.~D. Bruce:
\newblock \emph{J. Phys. A} \textbf{11}  (1978) 595

\bibitem{SG78}
J.~Sak, G.~S. Grest:
\newblock \emph{Phys. Rev. B} \textbf{17}  (1978) 3602

\bibitem{dAL}
L.~C. de~Albuquerque, M.~M. Leite:
\newblock \emph{J. Phys. A} \textbf{34}  (2001) L327;
\textbf{35}  (2002) 1807

\bibitem{DSZ}
H.~W. Diehl, M.~Shpot, R.~K.~P. Zia:
\newblock to be published

\end{thebibliography}

\end{document}